\documentclass[%
amsmath,amssymb,aps,showkeys, nofootinbib, twocolumn,superscriptaddress,longbibliography
]{revtex4-2}
\usepackage{graphicx}
\usepackage{float}
\graphicspath{{./figures/}}
\usepackage{dcolumn}
\usepackage{bm}
\usepackage{braket}
\usepackage{color}
\usepackage[normalem]{ulem}
\usepackage[colorlinks=true,linkcolor=blue,urlcolor=blue,citecolor=blue]{hyperref}

\begin{document}

\title{Chiral Magnon Mixing by Symmetry-Breaking in Collinear Ferrimagnets}

\author{Dhurba~R.~Jaishi}
\affiliation{Ames National Laboratory, Ames, IA, 50011, USA}
\affiliation{Department of Physics and Astronomy, Iowa State University, Ames, IA, 50011, USA}

\author{Tyler~J.~Slade}
\affiliation{Ames National Laboratory, Ames, IA, 50011, USA}
\affiliation{Department of Physics and Astronomy, Iowa State University, Ames, IA, 50011, USA}

\author{S.~X.~M.~Riberolles}
\affiliation{Ames National Laboratory, Ames, IA, 50011, USA}
\affiliation{Department of Physics and Astronomy, Iowa State University, Ames, IA, 50011, USA}

\author{Bing~Li}
\affiliation{Neutron Scattering Division, Oak Ridge National Laboratory, Oak Ridge, TN, 37831, USA}

\author{Tianxiong~Han}
\affiliation{Ames National Laboratory, Ames, IA, 50011, USA}
\affiliation{Department of Physics and Astronomy, Iowa State University, Ames, IA, 50011, USA}

\author{D.~M.~Pajerowski}
\affiliation{Neutron Scattering Division, Oak Ridge National Laboratory, Oak Ridge, TN, 37831, USA}

\author{D.~L.~Abernathy}
\affiliation{Neutron Scattering Division, Oak Ridge National Laboratory, Oak Ridge, TN, 37831, USA}

\author{Barry Winn}
\affiliation{Neutron Scattering Division, Oak Ridge National Laboratory, Oak Ridge, TN, 37831, USA}

\author{Melissa Graves-Brook}
\affiliation{Neutron Scattering Division, Oak Ridge National Laboratory, Oak Ridge, TN, 37831, USA}

\author{B.~G.~Ueland}
\affiliation{Ames National Laboratory, Ames, IA, 50011, USA}
\affiliation{Department of Physics and Astronomy, Iowa State University, Ames, IA, 50011, USA}

\author{R.~J.~McQueeney}
\affiliation{Ames National Laboratory, Ames, IA, 50011, USA}
\affiliation{Department of Physics and Astronomy, Iowa State University, Ames, IA, 50011, USA}

\date{\today}

\begin{abstract}
    Magnons in ferromagnets possess spin angular momentum defined by right-handed precession of the moment around the magnetization direction.  In antiferromagnets with no net magnetization, left- and right-handed magnons are degenerate in the absence of an applied field. Ferrimagnets possess uncompensated magnetic sublattices, which should natively possess right-and left-handed magnons where their energy is split by the internal molecular field.  Here, we show that $R$Mn$_6$Sn$_6$ ($R =$ Tb, Er) ferrimagnets possess right and left-handed magnon bands that cross at finite momentum (${\bf k}$) within the basal plane, defining modes with opposite dynamical chirality.  Depending on the symmetry of the ferrimagnetic order, which may be manipulated by varying the rare-earth magnetic anisotropy or with applied field, the band crossing may remain a nodal line or may be gapped. The gapped modes contain hybridized chiral excitations whose chirality becomes {\bf k}-dependent.
\end{abstract}

\maketitle

{\it Introduction.--}
    Magnons are quantized excitations that represent the collective precessional motion of ordered spins in magnetic solids.  As fundamental carriers of spin angular momentum (SAM, $\boldsymbol{\mathcal{S}}$) without net charge transport, magnons offer a platform for low-dissipation information transfer and processing. Magnon modes with right- or left-handed precession (opposite SAM) offer another means to encode and process information. Access to such modes provides a two-level basis set for representing binary information or constructing coherent superpositions (such as linearly polarized magnons) that open the opportunity to utilize magnons for quantum information processing~\cite{Daniels18}.  These properties form a foundation for spin-based logic, magnon transistors, and other elements of wave-based and neuromorphic computing~\cite{Lenk11,Chumak15,Pirro21,Yuan22}. When discussing propagation of magnon modes with crystal momentum ${\bf k}$, it is useful to define its intrinsic dynamical chirality as $\chi = \boldsymbol{\mathcal{S}} \cdot \hat{\mathbf{k}}/\hbar$, which is odd under reflection and even under time reversal.

    In ferromagnets, magnons exhibit only a right-handed precession about the magnetization direction and its mode chirality at ${\bf k}$ is single-valued. Unlocking a magnon’s chiral degree-of-freedom requires an additional magnetic sublattice. In many antiferromagnets, the symmetry between oppositely magnetized sublattices enforces the degeneracy of left- and right-handed modes, which can be lifted (for example) by the application of a magnetic field. On the other hand, the special symmetry properties of altermagnets allow opposite chirality magnons to be accessed in particular ${\bf k}$ directions without breaking global time-reversal symmetry~\cite{vsmejkal2023chiral,PhysRevLettMnTe}.  Ferrimagnets, in contrast, possess unequal magnetic sublattices that create an intrinsic molecular field, naturally splitting into branches of opposite chirality even in zero external field.  Clear observation of the opposite chirality of acoustic (right-handed) and optic (left-handed) ferrimagnetic magnons has been demonstrated by polarized inelastic neutron scattering (INS)~\cite{nambu2020observation}. 

    This built-in asymmetry and internal field of ferrimagnets provide a fertile setting for manipulating magnon chirality by reversing its handedness (SAM) at fixed ${\bf k}$.  In one example, the acoustic magnon chirality can be reversed when the material is tuned between its magnetic and angular momentum compensation temperature, allowing for selective excitation of right and left-handed magnons~\cite{kim2020distinct,mori2023magnetic,PhysRevLettWang,Yan25}. Similar chirality reversal through an external magnetic field, magnon-magnon coupling mediated by dipolar interactions, and interlayer exchange have also been observed in low-dimensional engineered systems~\cite{Shiota24,zhang2025switchable,PhysRevBLi,PhysRevBSud,liu2022, PhysRevLett2020, PhysRevLett2025, Duan25}. Here, we describe another mechanism where hybridization of chiral magnons may be induced by symmetry-breaking in the magnetic structure, allowing for the manipulation of magnon chirality.

    We investigate the chiral properties of magnons in the $R$Mn$_6$Sn$_6$ ferrimagnets ($R$166, where $R=$ rare-earth). More recently known for their topological electronic properties~\cite{yin2020quantum, ma2021rare, dhakal2021, Lee2023}, $R$166 magnets were originally investigated for their rich magnetic properties associated with the magnetic anisotropy of $R$ ions and competing $R$-Mn and Mn-Mn magnetic interactions~\cite{venturini1991, venturini1996, malaman1999,rosenfeld, PhysRevX_SXM, Riberolles24, kyle2024}. 
    
\begin{figure*}[ht]
        \centering
        \includegraphics[width=1.0\linewidth]{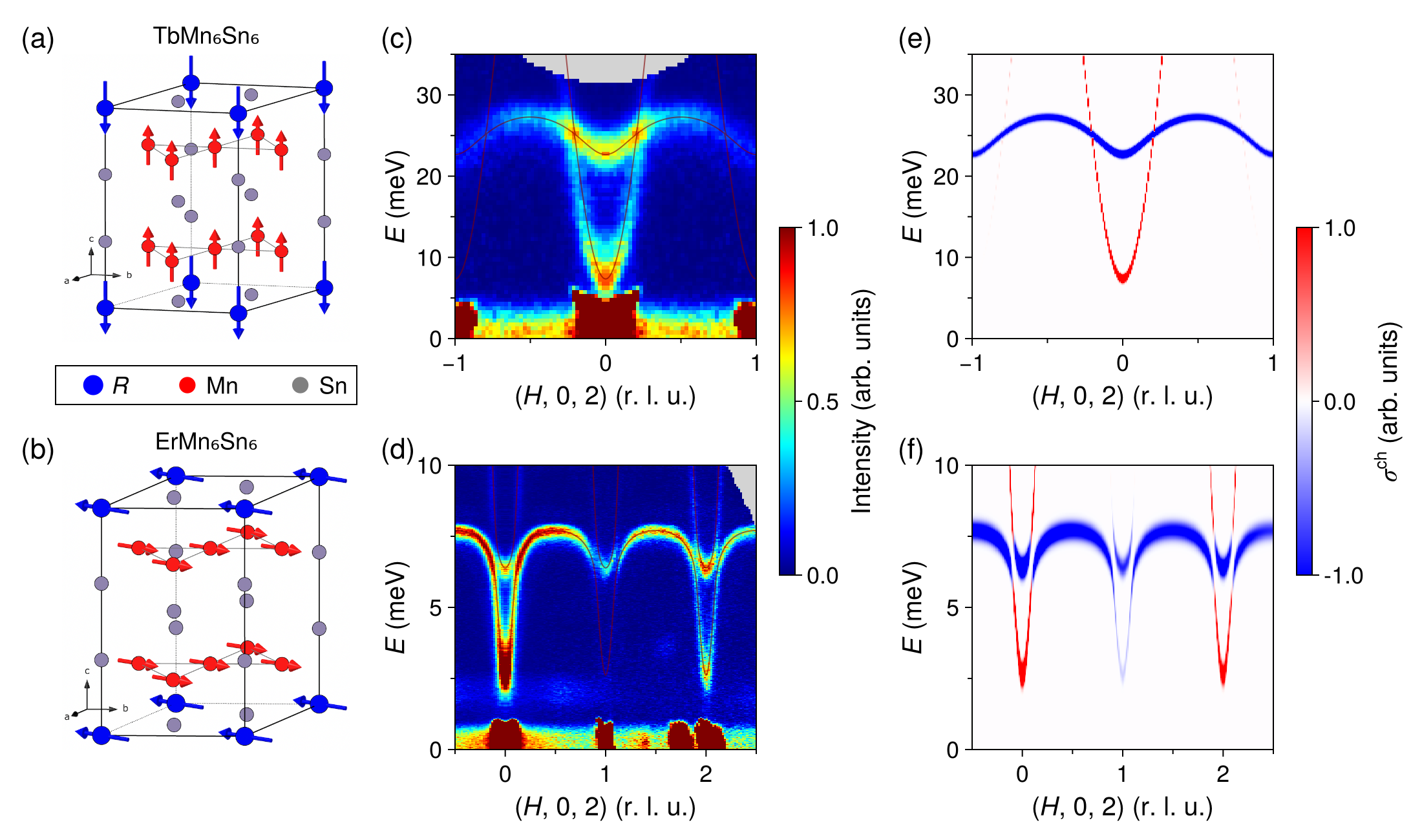}
        \caption{{\bf  Chiral magnons in Tb166 and Er166}. (a) Uniaxial ferrimagnetic structure of Tb166. (b) Planar ferrimagnetic structure of Er166. (c), (d) Unpolarized INS measurement of the magnon dispersions of Tb166 and Er166, respectively. For Tb166, measurements were performed with $E_i$ = 75 meV at 6 K along the reciprocal-lattice direction ($H$, 0, 2). Data have been integrated in the transverse directions with $K$ = [-0.05, 0.05] and $L$ = [1.90, 2.10] in reciprocal lattice units (r. l. u.). For Er166, measurements were performed with $E_i$ = 12 meV at 1.8 K along the reciprocal-lattice direction ($H$, 0, 2). Data have been integrated in the transverse directions with $K$ = [-0.03, 0.03] and $L$ = [1.80, 2.20] (r. l. u.). In (c) and (d), the solid lines correspond to the model dispersion based on parameters from Refs.~\cite{PhysRevX_SXM} and~\cite{Riberolles24}, respectively. (e), (f), Calculated chiral scattering of magnon modes in Tb166 and Er166, respectively, using linear spin wave theory and convoluted with a Gaussian energy full-width-at-half-maximum of 0.4 meV.}
        \label{fig:sqw_chi_R166}
    \end{figure*}

    The high-symmetry uniaxial ferrimagnetic structure of TbMn$_6$Sn$_6$ shown in Fig.~\ref{fig:sqw_chi_R166}(a) preserves $\text{U(1)}$ symmetry for magnon mode precessions. The circularly polarized right-handed acoustic and left-handed optic magnon bands are eigenstates of the SAM with quantized values of $\mathcal{S}_z=\mp \hbar$. Our INS measurements reveal a key feature in $R$166 compounds where acoustic and optic magnon bands cross at finite momentum {\bf k}.  In Tb166, the crossing is protected by $\text{U(1)}$ symmetry, and the SAM sectors do not mix.

    ErMn$_6$Sn$_6$ has a planar ferrimagnetic structure shown in Fig.~\ref{fig:sqw_chi_R166}(b) where magnetic anisotropy breaks $\text{U(1)}$ symmetry. The magnon SAM is no longer a good quantum number, and magnon polarizations are superpositions of the right and left-handed SAM basis states, resulting in elliptically polarized acoustic and optic magnons. This mixing allows for a hybridization to occur at the magnon band crossing, leading to the reversal of the magnon SAM and chirality within each band. Our polarized INS measurements directly measure the magnon SAM and confirm this reversal. Given the straightforward capability to access spin-reorientation transitions between high and low symmetry ferrimagnetic states in $R$166 compounds~\cite{malaman1999,riberolles2023orbital,Nil25}, this discovery generates an opportunity for facile magnon chirality switching with temperature or applied fields.
    \begin{figure*}[ht]
        \centering
        \includegraphics[width=1.0\linewidth]{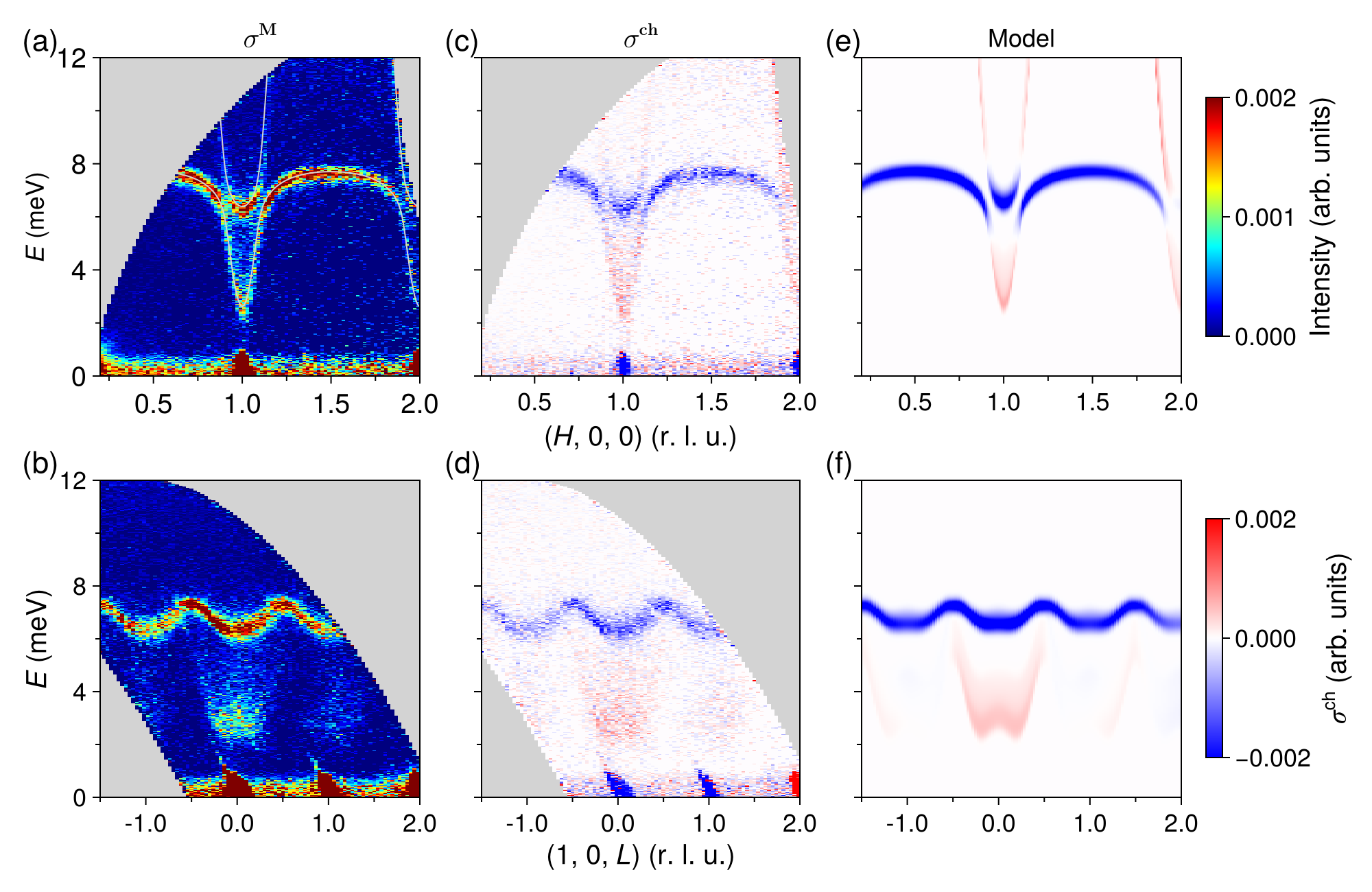}
        \caption{{\bf Magnon chirality mixing in Er166}.  Half-polarized INS data for Er166 showing the magnetic contribution ($\sigma^{\text{M}}$) (a)--(b) and the chiral contribution (c)--(d) to spin excitations along the reciprocal-lattice direction ($H$, 0, 0) and (1, 0, $L$), respectively. The scattering events were collected with $E_i$ = 15 meV at 6 K.  The gray solid lines in (a) correspond to the model dispersion. (e), (f) Model calculations of the chiral scattering along the two directions convoluted with a Gaussian energy full-width-at-half-maximum of 0.4 meV. The gray shaded regions in (a-d) are obscured by the kinematic limit.}
        \label{fig:sqw_chi_hys}
    \end{figure*}
    %

{\it Experimental results.--}
    To measure the magnon modes and their SAM, we performed half-polarized and unpolarized INS experiments on Tb166 and Er166. Fig.~\ref{fig:sqw_chi_R166}(c) and (d) show unpolarized data for Tb166 and Er166 measured respectively on the ARCS and CNCS spectrometers at the Spallation Neutron Source (SNS) at Oak Ridge National Laboratory. Experimental details are provided in the Supplementary Information (SI)~\cite{sup}, and Tb166 data were previously reported in Ref.~\cite{PhysRevX_SXM}. Each spectrum contains an acoustic and optic magnon mode along the ($H$, 0, 2) direction within the hexagonal Brillouin zone. In Tb166, the two modes cross and do not mix, but it is clear that the modes hybridize and repel in Er166.

    We can learn more about the character of these modes by calculating the chiral term in the magnetic cross-section for neutrons using linear spin-wave theory based on magnetic models developed in Refs.~\cite{PhysRevX_SXM,Riberolles24,Nil25}. The chiral scattering cross-section at momentum transfer ${\bf Q}$ is given by
    \begin{align}
         &\sigma^{ch}({\bf Q},\omega)  \nonumber \\
         &\propto\sum_{j,k}\text{e}^{i\textbf{Q}
        \cdot(\textbf{r}_k-\textbf{r}_j)} f^\ast_j({\bf Q}) f_k({\bf Q}) \nonumber \\
        &\qquad \times \frac{1}{2\pi \hslash}\int^{\infty}_{-\infty}
        i\big\langle M^{\alpha\dagger}_{\perp j}(0) M^{\beta}_{\perp k}(t) - M^{\beta\dagger}_{\perp j}(0)  M^{\alpha}_{\perp k}(t)\big\rangle \text{e}^{-i\omega t}dt \nonumber \\
        & \propto i[\langle{M}_{\perp}^{\alpha \dagger}({\bf Q}) M_{\perp }^{\beta}({\bf Q},t)\rangle_\omega - \langle{M}_{\perp}^{\beta \dagger}({\bf Q})M_{\perp}^{\alpha}({\bf Q},t)\rangle_\omega].
    \label{eqn:chiral_term}    
    \end{align}
    Here, $M^{\alpha(\beta)}_{\perp}$ are the components of the dynamical magnetization vector perpendicular to {\bf Q} for spins located at ${\bf r}_j$ and ${\bf r}_k$. $f_j({\bf Q})$ is the magnetic form factor. For excitations, the ``chiral" term measures antisymmetric spin-spin correlations and is related to the SAM of the mode weighted by the magnetic structure factor. The intrinsic dynamical chirality of a mode can be obtained by determining the projection of SAM along the momentum ${\bf k}$ (See SI~\cite{sup}).

    The calculated chiral scattering for Tb166 is shown in Fig.~\ref{fig:sqw_chi_R166}(e). Here, the positive (red) and negative (blue) values of the chiral scattering intensity distinguish the handedness of the magnon modes. The red color corresponds to counter-clockwise precession with respect to magnetization direction (right-handed polarization), whereas the mode with blue color corresponds to clockwise precession (left-handed polarization). We identify the steep mode as a right-handed (Mn-dominant) acoustic mode and the flatter mode as left-handed (Tb-dominant) optic mode.  In the case of Er166, calculations in Fig.~\ref{fig:sqw_chi_R166}(f) indicate that the lower branch is right-handed near $H=0$, but is predicted to reverse close to the hybridization gap and become left-handed, suggesting a mixing and reversal of the mode's dynamical chirality.  The opposite is true for the upper branch.  We note that both modes in Fig.~\ref{fig:sqw_chi_R166}(f) appear to be left-handed near $H=1$, but this arises from a reduced contribution of the Mn structure factor to the chiral cross-section in this Brillouin zone.
    
    Half-polarized neutron scattering measurements performed on Er166 using the HYSPEC instrument at SNS confirm the chiral magnon mixing and reversal predicted by our calculations (see~SI~\cite{sup}). Measurements were performed in an applied field of ${\bf B}=$ 0.12 Tesla along the planar easy-axis ${\bf a}^*$, sufficient to polarize ferrimagnetic domains and provide a uniform net magnetization ${\bf M}_0 \parallel {\bf a}^*$. The neutron polarization of the incident beam ({\bf P}) is defined to be either parallel or antiparallel to both {\bf B} and {\bf Q}; ${\bf P}\parallel{\bf Q}\parallel{\bf B}$.  The scattered neutrons are recorded in two channels  $I^{+ 0}$ and $I^{- 0}$ ( shown in SI Fig.~S2~\cite{sup}) where $+$ and $-$ indicate the polarization of the incident neutrons. The flipping ratio of 15 measured at (1, 0, 0) Bragg peak for $E_i$ of 15 meV, indicating minimal depolarization of the incident beam. The chiral spectra from Eqn.~(\ref{eqn:chiral_term}) are extracted from the channel difference, $\sigma^{\text{ch}} = \frac{1}{2}(I^{+0}-I^{-0})$~\cite{chatterji2006neutron,simonet2012magnetic}. The channel average measures the conventional magnetic response for unpolarized neutrons, $\sigma^{\text{M}} = \frac{1}{2}(I^{+0}+I^{-0}) \propto \langle{M}_{\perp}^{\alpha}(Q) {M^\dagger}_{\perp}^{\alpha}(Q,t)\rangle_\omega$.
     
    Figures~\ref{fig:sqw_chi_hys}(a)--(b), show the magnetic dispersion ($\sigma^\text{M}$) in the in-plane and out-of-plane direction, respectively, with the former providing equivalent magnon spectra to the unpolarized INS data shown in Fig.~\ref{fig:sqw_chi_R166}(d). The chiral scattering ($\sigma^\text{ch}$) is shown in Figs.~\ref{fig:sqw_chi_hys}(c)--(d) and confirms the predicted reversal of the magnon handedness close to the hybridization point. Calculations of the chiral scattering shown in Figs.~\ref{fig:sqw_chi_hys}(e) and (f) are in good agreement with observations.

{\it Magnetic symmetry.--}
    We have discovered that magnon hybridization and chiral mixing will occur in Er166, but not in Tb166. Whether or not hybridization occurs has to do with the different symmetry of the planar and uniaxial ferrimagnetic structures. The key to chiral magnon mixing is the presence or absence of $\text{U(1)}$ symmetry for phase rotations around the magnetization direction (precessional motion). Easy-plane anisotropy for phase rotations conserves $\text{U(1)}$ symmetry, leading to circular precession of moments in a magnon mode. The right and left-handed circular magnon polarizations ($\boldsymbol{\mu}$) have quantized values of the magnon SAM ($\mathcal{S}_z = \mp \hbar$) where $\boldsymbol{\mathcal{S}}=i\hbar(\boldsymbol{\mu}^*\times\boldsymbol{\mu})$. The SAM basis eigenstates can be represented as 
    \begin{equation}
        \ket{\pm} = \ket{S^{\alpha}} \pm i\ket{S^{\beta}}
    \end{equation}
    where $S^{\alpha}$ and $S^{\beta}$ are spin components that are perpendicular to the static moment direction.  
    
    For Tb166, the uniaxial ferrimagnetic order with moments along $z$ is shown Fig.~\ref{fig:sqw_chi_R166}(a). While Tb166 strictly has sixfold symmetry for phase rotations, anisotropic effects related to $B^6_6$ crystal-field terms are sixth-order in the spin operators and will vanish within linear spin-wave theory, maintaining effective $\text{U(1)}$ symmetry. The right and left-handed polarizations in Tb166 are therefore diagonal in the SAM basis; $\ket{R} = \ket{+}$ and $\ket{L} = \ket{-}$, respectively.  There is no term ($V$) in the spin-wave Hamiltonian that will mix the two SAM sectors ($\bra{R}V\ket{L}=\bra{+}V\ket{-}=0$).  Thus, the SAM and chirality are good quantum numbers that are protected by $\text{U(1)}$ symmetry.

    For the Er166 planar ferrimagnetic structure shown in Fig.~\ref{fig:sqw_chi_R166}(b), there is twofold rotational symmetry around the magnetization. The presence of Er and Mn polar magnetic anisotropy energies of the form $V=K_{yy}S_{y}^2 + K_{zz}S_z^2$ will break $\text{U(1)}$ symmetry.  This generally leads to right and left-handed elliptical polarizations that are linear superpositions of the SAM eigenstates.
    \begin{equation}
        \ket{R,L} = c_{+}\ket{+} + c_{-}\ket{-}
    \end{equation}
    Here, $\ket{R}$ ($\ket{L}$) occurs when $c_+>c_-$ ($c_+<c_-$), respectively, and generally the SAM of the magnon mode is not conserved. The interaction between modes of different handedness is allowed, $\bra{R}V\ket{L}\neq0$, since there are off-diagonal matrix elements $\bra{\pm}V\ket{\mp} \propto |K_{yy}- K_{zz}|$ that couple the two SAM states. Thus, mode hybridization occurs with a gap proportional to the net $yz$ anisotropy.
    
{\it Mode chiralities.--}
    While the neutron scattering data are definitive in establishing the chirality reversal in Er166, the chiral term in the neutron intensity in Eq.~(\ref{eqn:chiral_term}) is weighted by both the magnon eigenvector and the structure factor.  Thus, the sign of the chiral scattering for a given mode can change in different Brillouin zones, as shown in Fig.~\ref{fig:sqw_chi_R166}(f). The mode SAM can be obtained from evaluations of the magnon eigenvectors based on an accurate model Hamiltonian using linear spin-wave theory. 
    \begin{figure}[ht]
        \centering
        \includegraphics[width=1.0\columnwidth]{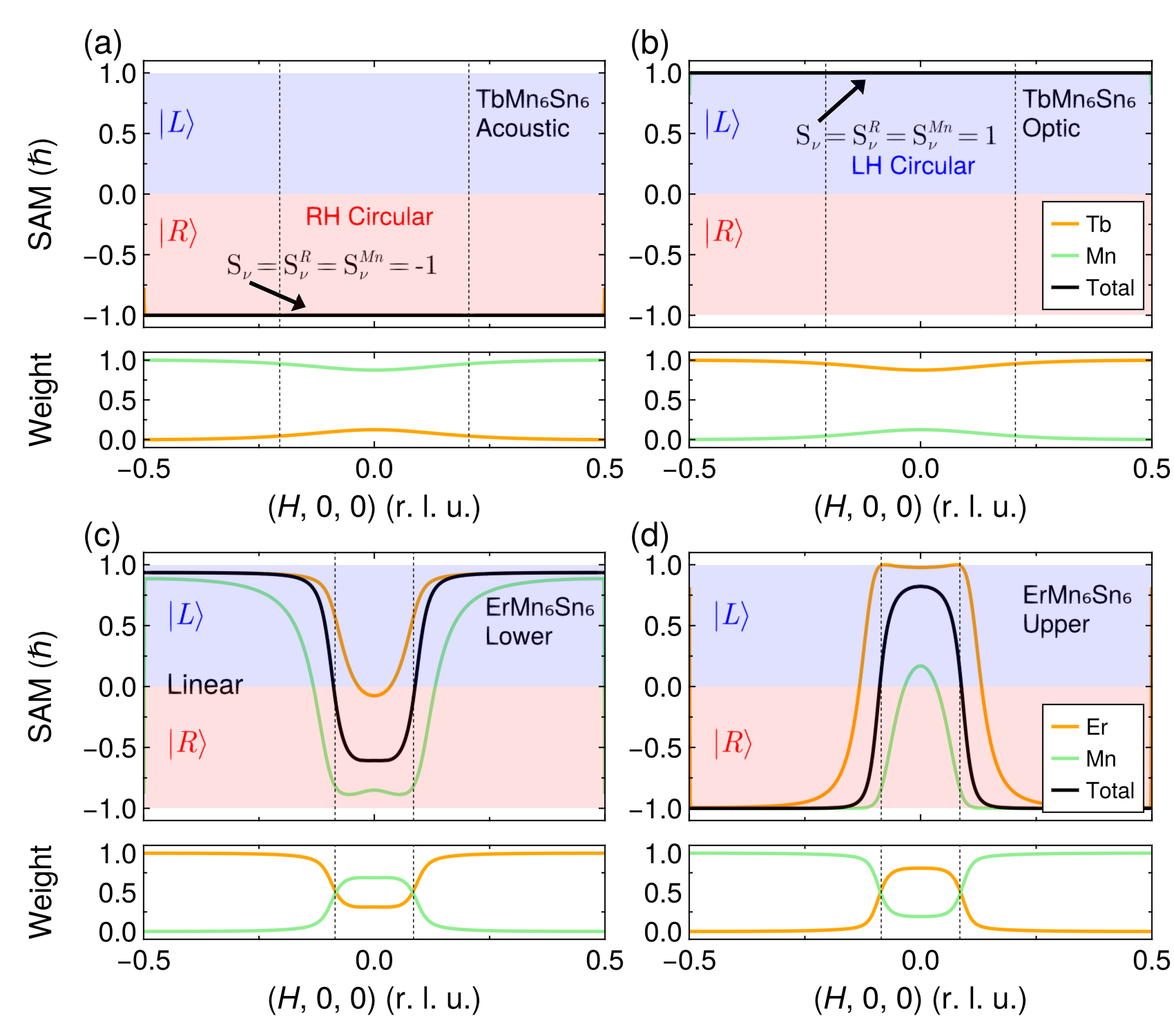}
        \caption{{\bf Magnon spin angular momentum}. Sublattice SAM and total SAM (in units of $\hbar$) for acoustic and optic modes in (a), (b) Tb166 and (c), (d) Er166. In each panel, the lower plot shows the sublattice weight for each eigenvector.  Thin vertical lines indicate the wavevector where the modes cross.}
    \label{fig:mode_chirality}
    \end{figure}
    The normalized magnon mode $\nu$ describes the precession of spin $j$ in the plane perpendicular to the net magnetization direction $\hat{\bf{M}}_0$,
    \begin{equation}
        \boldsymbol{\mu}_{\nu j} = a_{\nu j}\hat{\boldsymbol{\epsilon}}_{\nu j}^{1} \pm ib_{\nu j}\hat{\boldsymbol{\epsilon}}_{\nu j}^2
    \end{equation}
    Here $\hat{\boldsymbol{\epsilon}}_{\nu j}^n$ describe the axes of elliptical precession of spin $j$ with eccentricity $\sqrt{1-a_{\nu j}^2/b_{\nu j}^2}$ and a SAM along the magnetization direction of $\mp2\hbar a_{\nu j}b_{\nu j}$.
    
    The total SAM of a magnon mode is given by the sum over the $R$ and Mn sublattices
    \begin{equation}
        \mathcal{S}_{\nu} = W_{\nu Mn}\mathcal{S}_{\nu Mn} + W_{\nu R}\mathcal{S}_{\nu R}
    \label{eqn:total_chirality}
    \end{equation}
    where $\mathcal{S}_{\nu r}$ is the sublattice SAM and $W_{\nu r}$ is the weight of sublattice $r = R$ or Mn in mode $\nu$. The sublattice SAM is defined as
    \begin{equation}
        \mathcal{S}_{\nu r} = i\hbar\hat{\bf M}_0\cdot \sum_{j \in r}  (\boldsymbol{\mu}_{\nu j}^* \times \boldsymbol{\mu}_{\nu j})/\sum_{j \in r} |\mu_{\nu j}|^2
    \label{eqn:sublattice_chirality}
    \end{equation}
    and the sublattice weight of mode $\nu$ is
    \begin{equation}
        W_{\nu r} = \sum_{j \in r} |\mu_{\nu j}|^2/\sum_{j} |\mu_{\nu j}|^2
    \end{equation}
    such that $W_{\nu R}+W_{\nu Mn}=1$. The dynamical chirality of the magnon is well-defined and proportional to the SAM when the finite momentum ${\bf k}$ lies along the magnetization direction ($\chi_{\nu }=\mathcal{S}_{\nu}/\hbar$).
    
\begin{figure}[!h]
        \centering
        \includegraphics[width=1.0\columnwidth]{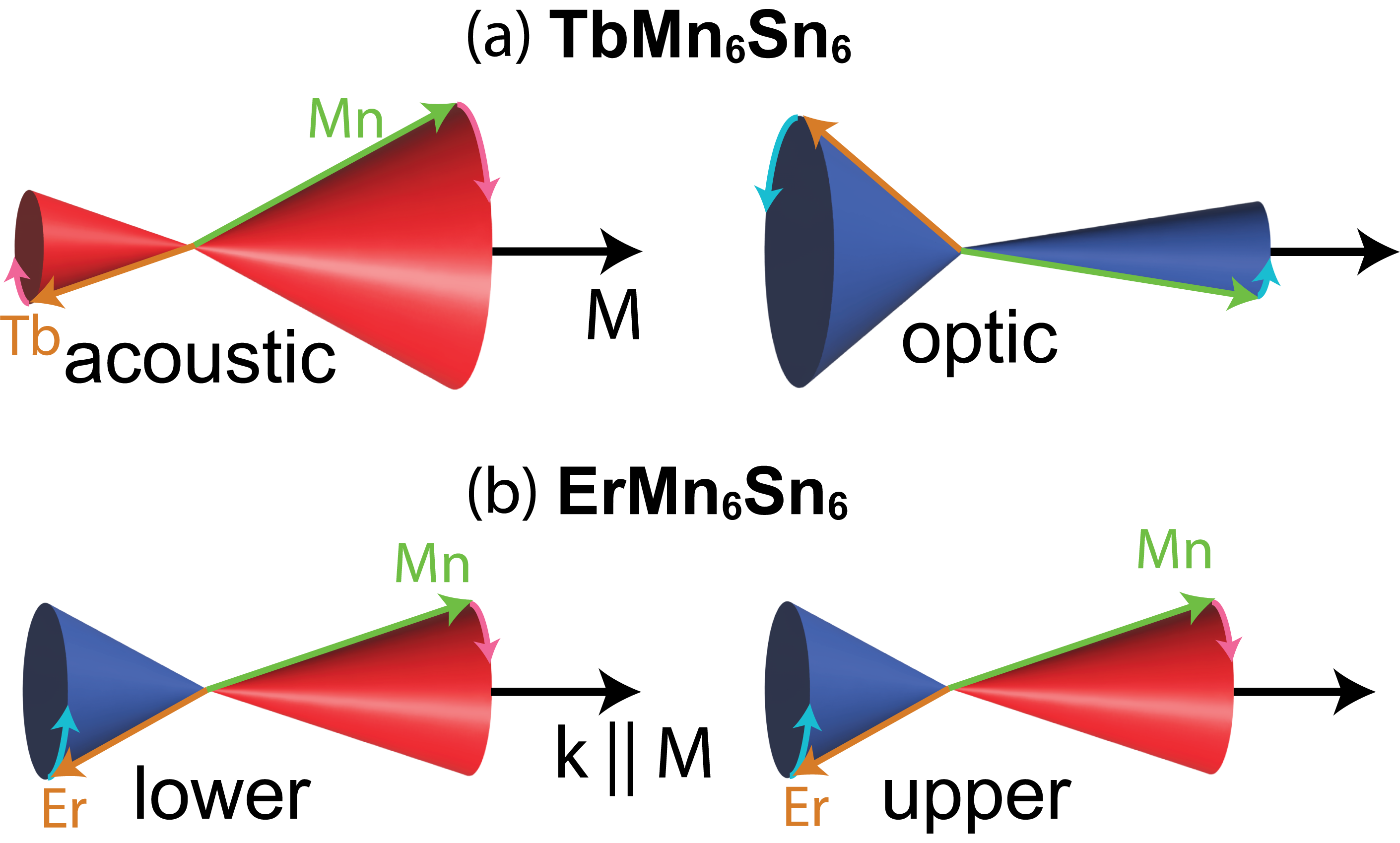} 
        \caption{{\bf Magnon mode eigenvectors.} Precession of rare-earth moments (orange arrow) and Mn moments (green arrow) in a right-handed (red cone) or left-handed fashion (blue cone) relative to the net magnetization ({\bf M}).  Eigenvectors are shown near the mode-crossing wavevector for (a) Tb166 and (b) Er166.  For Er166, the Er and Mn moments precess with opposite handedness, allowing the cancellation of the net SAM and chirality.}
    \label{fig:result_magnon}
    \end{figure}

    Figs.~\ref{fig:mode_chirality}(a) and (b) show calculations of the SAM for the acoustic and optic modes of Tb166. The acoustic mode is right-handed, and the weight is dominated by Mn, while the optic mode is left-handed and dominated by Tb.  Both modes are circularly polarized ($\mathcal{S}_{\nu} = \mp \hbar$) and each sublattice has the same handedness, indicating modes of pure SAM basis confirming the proposition above that $\text{U(1)}$ symmetry is conserved for Tb166. Magnon polarizations are shown schematically in Fig.~\ref{fig:result_magnon}(a).
    For Er166, the two modes hybridize, leading to a lower and upper branch.  Near $H=0$, Fig.~\ref{fig:mode_chirality}(c) shows that the lower branch is right-handed and each sublattice has right-handed elliptical polarization ($|\mathcal{S}_{\nu}| < \hbar$) such that the chirality is not pure.  As $H$ gets larger, there is a point where the total chirality is zero and then reverses in the lower branch.  In the case of photons, a zero-chirality state would signify linear polarization. However, in this case, we see that the Mn and $R$ sublattices are still elliptically polarized, but now with opposite handedness, which leads to cancellation of the SAM as shown in Fig.~\ref{fig:mode_chirality}(c). There is also a transfer of the dominant weight of the lower branch from Mn to Er as the modes cross. Before the hybridization occurs, there is a point where the Er sublattice itself reverses from right to left-handed elliptical polarization and is linearly polarized at this point, which is uncommon for magnons. The upper branch has a similar evolution with the mode weights and chiralities reversed.
    
    We report the observation of a novel mechanism where magnetic symmetry and induced anisotropy control the mixing of magnon chiralities in a ferrimagnet. In this respect, the hybridized chiral modes resemble magnon squeezing effects, which are driven by coherent coupling between magnon modes mediated by anisotropic interactions~\cite{kam2017,lukas2019,kam2020}. The mixing results in the reversal of the chirality with momentum within a single band. The symmetry dependence can allow for control of magnon chirality when magnetic fields or temperature drive to metamagnetic (spin-reorientation) transitions (See SI~\cite{sup}). Our observation provides a new mechanism for advancing the chirality-based quantum magnonics.

    {\it Acknowledgements.--} 
    We thank Andrei~T.~Savici and Vasile~O.~Garlea for valuable discussions during the polarized INS experiment. We would also like to thank Thais V. Trevisan for thoughtful discussions and proofreading of the manuscript. This work is supported by the U.S. Department of Energy (U.S. DOE), Office of Basic Energy Sciences (BES), Division of Materials Sciences and Engineering through the Ames National Laboratory under Contract No. DE-AC02-07CH11358. TJS synthesized and characterized crystals with funding from the Center for the Advancement of Topological Semimetals (CATS), an Energy Frontier Research Center funded by the U.S. DOE, BES, through the Ames National Laboratory. This research used resources at the Spallation Neutron Source, a DOE Office of Science User Facility operated by the Oak Ridge National Laboratory. The beam time was allocated to ARCS, CNCS, and HYSPEC on proposal numbers IPTS-26429.1,  IPTS-30676.1, and IPTS-34245.1, respectively.

    {\it Data availability.--} 
    The data that support the findings of this article are openly available~\cite{draj_data}.

\bibliography{reference.bib}
\end{document}


\title{Supplemental Materials \\ for \\ Chiral Magnon Mixing by Symmetry-Breaking in Collinear Ferrimagnets}

\author{Dhurba~R.~Jaishi}
\affiliation{Ames National Laboratory, Ames, IA, 50011, USA}
\affiliation{Department of Physics and Astronomy, Iowa State University, Ames, IA, 50011, USA}

\author{Tyler~J.~Slade}
\affiliation{Ames National Laboratory, Ames, IA, 50011, USA}
\affiliation{Department of Physics and Astronomy, Iowa State University, Ames, IA, 50011, USA}

\author{S.~X.~M.~Riberolles}
\affiliation{Ames National Laboratory, Ames, IA, 50011, USA}
\affiliation{Department of Physics and Astronomy, Iowa State University, Ames, IA, 50011, USA}

\author{Bing~Li}
\affiliation{Neutron Scattering Division, Oak Ridge National Laboratory, Oak Ridge, TN, 37831, USA}

\author{Tianxiong~Han}
\affiliation{Ames National Laboratory, Ames, IA, 50011, USA}
\affiliation{Department of Physics and Astronomy, Iowa State University, Ames, IA, 50011, USA}

\author{D.~M.~Pajerowski}
\affiliation{Neutron Scattering Division, Oak Ridge National Laboratory, Oak Ridge, TN, 37831, USA}

\author{D.~L.~Abernathy}
\affiliation{Neutron Scattering Division, Oak Ridge National Laboratory, Oak Ridge, TN, 37831, USA}

\author{Barry Winn}
\affiliation{Neutron Scattering Division, Oak Ridge National Laboratory, Oak Ridge, TN, 37831, USA}

\author{Melissa Graves-Brook}
\affiliation{Neutron Scattering Division, Oak Ridge National Laboratory, Oak Ridge, TN, 37831, USA}

\author{B.~G.~Ueland}
\affiliation{Ames National Laboratory, Ames, IA, 50011, USA}
\affiliation{Department of Physics and Astronomy, Iowa State University, Ames, IA, 50011, USA}

\author{R.~J.~McQueeney}
\affiliation{Ames National Laboratory, Ames, IA, 50011, USA}
\affiliation{Department of Physics and Astronomy, Iowa State University, Ames, IA, 50011, USA}

\maketitle

\onecolumngrid

\section{Experimental Methods}
\subsection{Polarized inelastic neutron scattering}
    Polarized inelastic neutron scattering (INS) measurements on ErMn$_6$Sn$_6$ were performed on the Hybrid Spectrometer (HYSPEC)~\cite{winn2015recent,Zaliznyak_2017} at the Spallation Neutron Source (SNS), Oak Ridge National Laboratory (ORNL). The incident beam is polarized by the vertically focusing Heusler crystal array (horizontally magnetized). The Mezei spin-flipper in front of the sample then controls the neutron polarization direction [spin-up (+) or spin-down (-)]. The neutron polarization at the sample is determined by using the adiabatic guide field provided by the coil system positioned at the sample table. In a half-polarization mode, the guide field along the neutron flight path between the polarizer and the sample maintains the neutron polarization. The scattered neutrons are recorded without performing polarization analysis; instead of a supermirror-analyzer array, a radial collimator is used.
    
\begin{figure}[!h]
        \centering
        \includegraphics[width=0.95\linewidth]{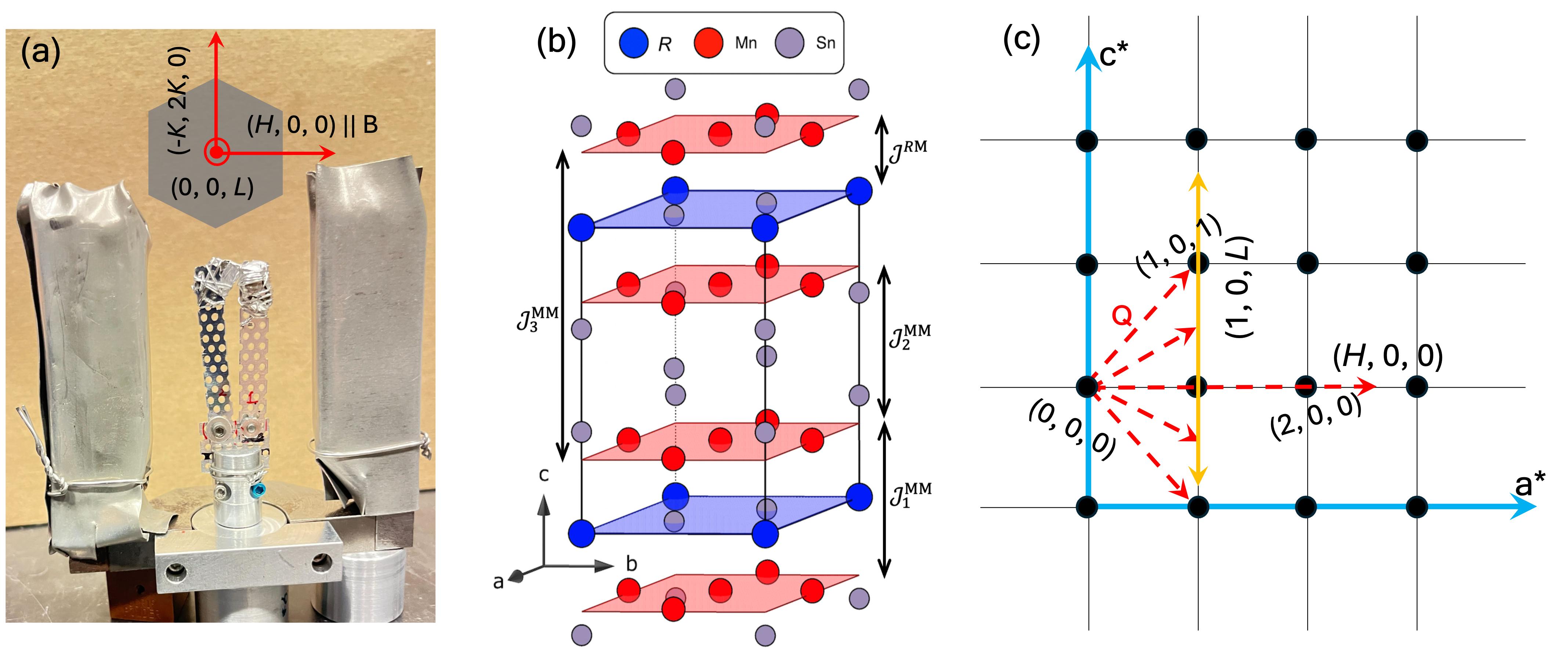}
        \caption{(a) Co-aligned single crystals in ($H$, 0, $L$) scattering plane, mounted with a magnetic yoke system. The inset shows the reciprocal-lattice direction. The magnetic field generated by magnets is parallel to ${\bf a}^*$($H$, 0, 0). The magnetic yoke is covered with cadmium to avoid a spurious background. (b) Crystal structure of $R$Mn$_6$Sn$_6$ with unit cell indicated by thin lines. The Mn--kagome ($R$--triangular) planes are indicated in red (red). The interlayer exchange interactions between Mn layers ($\mathcal{J}^{\text{MM}}_k$) and $R$--Mn layers ($\mathcal{J}^{\text{$R$M}}$) are shown by black double arrows. (c) Reciprocal lattice directions ($H$, 0, 0) and  (1, 0, $L$), where we measured scattering events. Dashed lines with an arrowhead represent {\bf Q} directions, and a yellow line with double arrows represents the scan along (1, 0, $L$).}
        \label{fig:cry_mag_yoke}
    \end{figure}

    The co-aligned single crystals [see Ref.~\cite{Riberolles24} for single crystal growth and characterization] with a total mass of $\approx$ 2.7 grams were aligned in the ($H$, 0, $L$) horizontal scattering plane using an aluminum holder as shown in  Fig.~\ref{fig:cry_mag_yoke}(a). To eliminate magnetic domains, we used a horizontal magnetic field parallel to the $(H, 0, 0)$ direction at the sample using two permanent magnets (Nd$_2$Fe$_{14}$B), one on each side of the sample by using a magnetic yoke and also additional guide field component in the vertical direction ($\perp {\bf a}^*$) of 20 G at the sample to eliminate the depolarization of scattered neutrons~\cite{nambu2020observation}.
    The permanent magnets with dimensions of $0.5''\times0.5''\times3''$ provided a field of $\approx$1200 G at the sample. The magnets and yoke system were fixed to the sample mount so that the field direction remained fixed with respect to the sample. The co-aligned samples and magnetic yoke system were mounted to a closed-cycle refrigerator. Because of a permanent magnet assembly to generate a magnetic field, field-reversal measurements were not performed. The data were collected at 6 K using an incident energy of $E_i$ = 15 meV with a Fermi chopper frequency of 180 Hz (elastic resolution full-width-at-half-maximum of 0.75 meV). We measured a flipping ratio of 15 for this $E_i$, which indicates minimal depolarization of the incident beam. The detector bank was positioned to provide a total horizontal scattering angle range of - 5$^{\circ}$  to - 85$^{\circ}$ to cover {\bf Q}, $E$ space, where {\bf Q} ($E$) is the momentum (energy) transfer, respectively.

\begin{figure}[!h]
        \centering
        \includegraphics[width=0.8\linewidth]{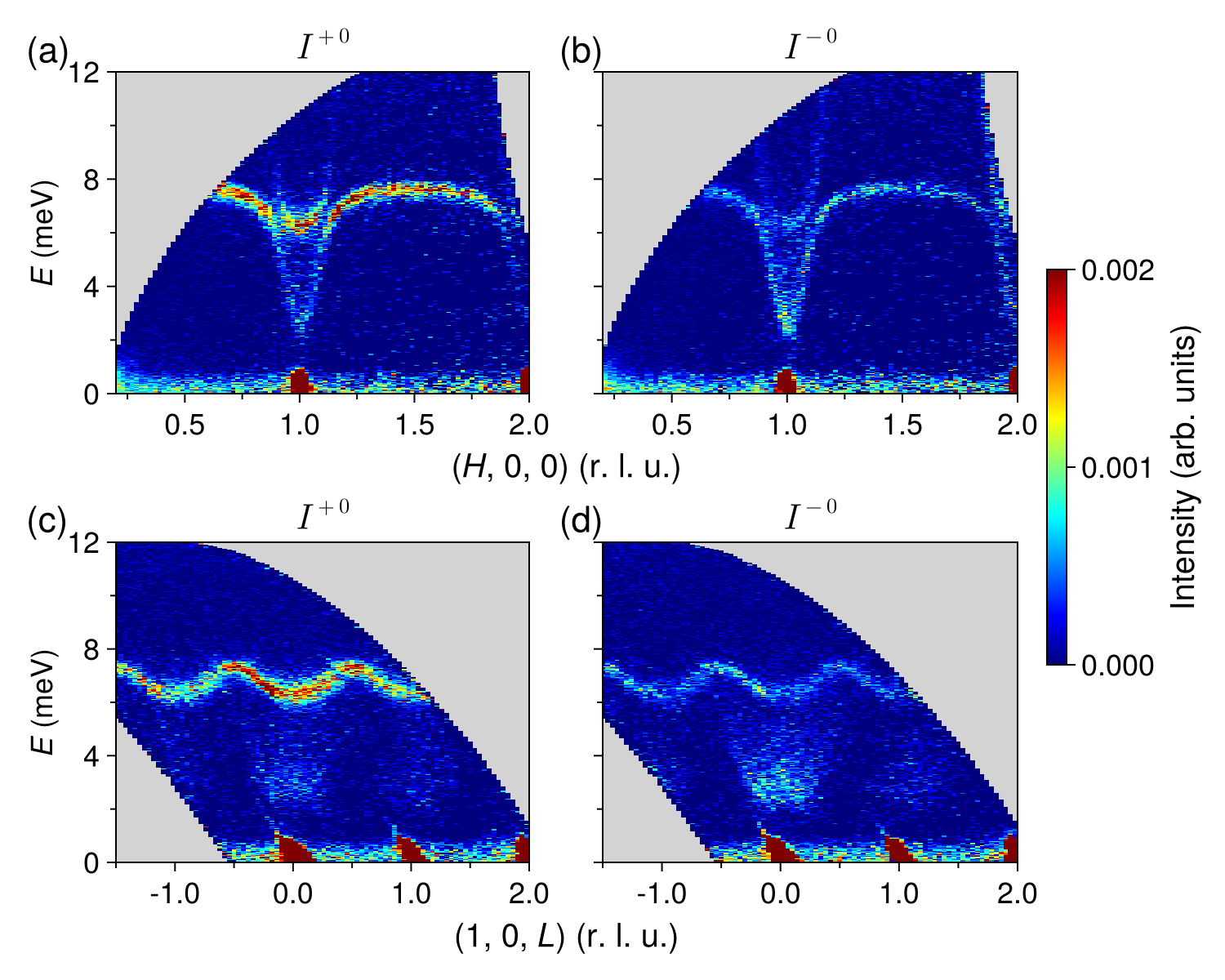}
        \caption{Spin excitations measured with polarized neutrons along the reciprocal-lattice direction ($H$, 0, 0) ((a)--(b)) and (1, 0, $L$) ((c)--(d)) for $I^{+0}$ (flipper off) and $I^{-0}$ (flipper on) channels respectively. The scattering events were collected with $E_i$ = 15 meV at 6 K. The energy resolution, full-width-at-half-maximum at the elastic line and around 6 meV (near the hybridization gap) is about 0.75 and 0.4 meV, respectively. The gray shaded regions are obscured by the kinematic limit.}
        \label{fig:raw_data_hyspec}
    \end{figure}

    The polarized INS data were collected as a function of energy and momentum transfer in the hexagonal reciprocal lattice units, {\bf Q} = $H{\bf a}^* + K{\bf b}^* + L{\bf c}^*$ $\equiv$ ($H$, $K$, $L$), where {\bf a}$^*$, {\bf b}$^*$, and {\bf c}$^*$ are the primitive reciprocal lattice vectors. Data are presented using three orthogonal hexagonal reciprocal vectors ($H$, 0, 0), (-$K$, 2$K$, 0), and (0, 0, $L$). For the half-polarized mode, with only one flipper before the sample, we measured two scattering channels: $I^{\pm 0}$~(Fig.\ref{fig:raw_data_hyspec}). The data are integrated in the transverse directions with $K$ = [-0.05, 0.05] and $L$ = [-0.08, 0.08] (r. l. u.) for in-plane dispersion. Similarly, with $H$ = [0.95, 1.05] and $K$ = [-0.05, 0.05] (r. l. u.) for out-of-plane dispersion.

\section{Model calculations}
\subsection{Model Hamiltonian}
    The simplest Hamiltonian that explains the observed spin dynamics in $R$166 compounds is
    \begin{equation}
        \mathcal{H}=\mathcal{H}_{\text{ex}} + \mathcal{H}_{\text{aniso}}
    \end{equation}
    where the exchange Hamiltonian is given by 
    \begin{align}
        \mathcal{H}_{\text{ex}}=&\sum_{\langle i<j \rangle}\mathcal{J}^{\text{MM}}_0 \mathbf{S}_i \cdot \mathbf{S}_j+
        \sum_{k}\sum_{i<j}\mathcal{J}^{\text{MM}}_k \mathbf{S}_i \cdot \mathbf{S}_j \nonumber \\
        & + \mathcal{J}^{\text{$R$M}}\sum_{\langle i<j\rangle}\mathbf{S}_i \cdot \mathbf{J}_j
    \end{align}
    Here, $\mathcal{J}^{\text{MM}}_0$ and $\mathcal{J}^{\text{MM}}_k$ are intralayer and interlayer exchange interactions between Mn--Mn
    spins shown in Fig.~\ref{fig:cry_mag_yoke}(b). $k$ indexes the different interlayer interactions. $\mathcal{J}^{\text{$R$M}}$ is the exchange interaction between $R$ and Mn spins. The angular brackets $\langle ... \rangle$ denote nearest neighbor sites. $\mathcal{J}^{\text{MM}}_0$, $\mathcal{J}^{\text{MM}}_1$, and $\mathcal{J}^{\text{MM}}_2$ are ferromagnetic, whereas $\mathcal{J}^{\text{MM}}_3$ and $\mathcal{J}^{\text{$R$M}}$ are antiferromagnetic. 
        
    The anisotropy energy includes easy-plane single-ion anisotropy for Mn and the crystalline-electric field (CEF) potential of neighboring ions on the 4$f$ orbital states of $R$ and is given by
    \begin{equation}
        \mathcal{H}_{\text{aniso}}=\sum_{i, \text{Mn}} K^{\text{M}}(S^z_i)^2
        + \sum_{i, \text{$R$}} B_l^m \mathcal{O}_l^m({\bf J}_i)
    \end{equation}
    Here $K^\text{M}$ $>\text{0}$ indicates easy plane anisotropy for Mn. $B_l^m$ are the CEF parameters for the $R^{3+}$ and $\mathcal{O}_l^m$ are Steven operators. For hexagonal point group symmetry, allowed CEF parameters are $B^0_2$, $B^0_4$, $B^0_6$, and $B^6_6$. Here, we used $S$ = 1, $gS$ $\approx$ 2.17 $\mu_B$ for Mn and $J$ = 15/2, $g_JJ$ $\approx$ 9 $\mu_B$ for Er for the calculations. The spin waves are computed using the Holstein-Primakov formalism~\cite{PhysRev1940} in the linear approximation using the \textsc{sunny.jl}~\cite{Dahlbom2025}. The fitted exchange and CEF parameters are summarized in the Table~\ref{tab:Heisen}.
    \begin{table}[!h]
        \centering
        \renewcommand{\arraystretch}{1.5}
        \setlength{\tabcolsep}{7pt} 
        \caption{Heisenberg and crystal field parameters for TbMn$_6$Sn$_6$ and ErMn$_6$Sn$_6$ (in meV).}
        \label{tab:Heisen}
        \begin{tabular}{l|c|c}
            \hline\hline
            & TbMn$_6$Sn$_6$ & ErMn$_6$Sn$_6$ \\
            \hline
            $\mathcal{J}^{\text{MM}}_0$ & $-$28.80 & $-$28.80 \\
            $\mathcal{J}^{R\text{M}}$   &   1.10   &    0.315 \\
            $\mathcal{J}^{\text{MM}}_1$ & $-$4.40  & $-$6.57  \\
            $\mathcal{J}^{\text{MM}}_2$ & $-$19.20 & $-$19.20 \\
            $\mathcal{J}^{\text{MM}}_3$ &   1.80   &    3.219 \\
            $K^\text{M}$                &   0.88   &    0.20  \\
            $B^0_2$                     &  $-8.675 \times 10^{-3}$ &  0.012 \\
            $B^0_4$                     &  $-1.430 \times 10^{-3}$ & $-4.059 \times 10^{-4}$ \\
            $B^0_6$                     & -- & -- \\
            $B^6_6$                     & -- & $0.588 \times 10^{-5}$ \\
            \hline\hline
        \end{tabular}
    \end{table}

\subsection{Neutron scattering cross-section}
    The partial differential scattering cross-section can be expressed as
    \begin{align}
            \label{eqn:neutron}
        \frac{d^2\sigma}{d\Omega dE_f} = \frac{k_f}{k_i}\left(\frac{\gamma r_0}{2}\right)^2\sum_{j,k}\text{e}^{i\textbf{Q}
        \cdot(\textbf{r}_k-\textbf{r}_j)}g_j f^\ast_j({\bf Q})g_k f_k({\bf Q})\times \frac{1}{2\pi \hslash} \int^{\infty}_{-\infty}
        \left[\Big\langle\textbf{S}^\dagger_{\perp j} \cdot \textbf{S}_{\perp k} \Big\rangle \right. + \left.i \textbf{P}
        \cdot\Big\langle\textbf{S}^\dagger_{\perp j} \times \textbf{S}_{\perp k}\Big\rangle \right]\text{e}^{-i\omega t}dt
    \end{align}
    where $k_i$ and $k_f$ are the initial and final neutron wave numbers, $\gamma$ = 1.913, and $r_0$ is the classical radius of the electron. $f_j$({\bf Q}) is the magnetic form factor. $\mathbf{S}_{\perp j}$ is the projection of the spin located at ${\bf r}_j$ on the plane perpendicular to the scattering vector $\mathbf{Q}$. Here, we have ignored the Debye-Waller factor and nuclear scattering. The first correlation term inside the square bracket is purely magnetic;
    \begin{align}    
        \sigma^{\text{M}} =  \sigma^y+\sigma^z \propto \frac{1}{2}(I^{+0}+I^{-0})
        \label{eqn:pol_mag}
    \end{align}
    and the second term is chiral correlation and for given the neutron polarization {\bf P} parallel to $x$;
    \begin{align}    
        \sigma^{\text{ch}} = \sigma^{yz}-\sigma^{zy} \propto \frac{1}{2}(I^{+0}-I^{-0})
        \label{eqn:pol_ch}
    \end{align}
\subsection{Switching of hybridized chiral magnon}
\begin{figure}[!h]
        \centering
        \includegraphics[width=1.0\linewidth]{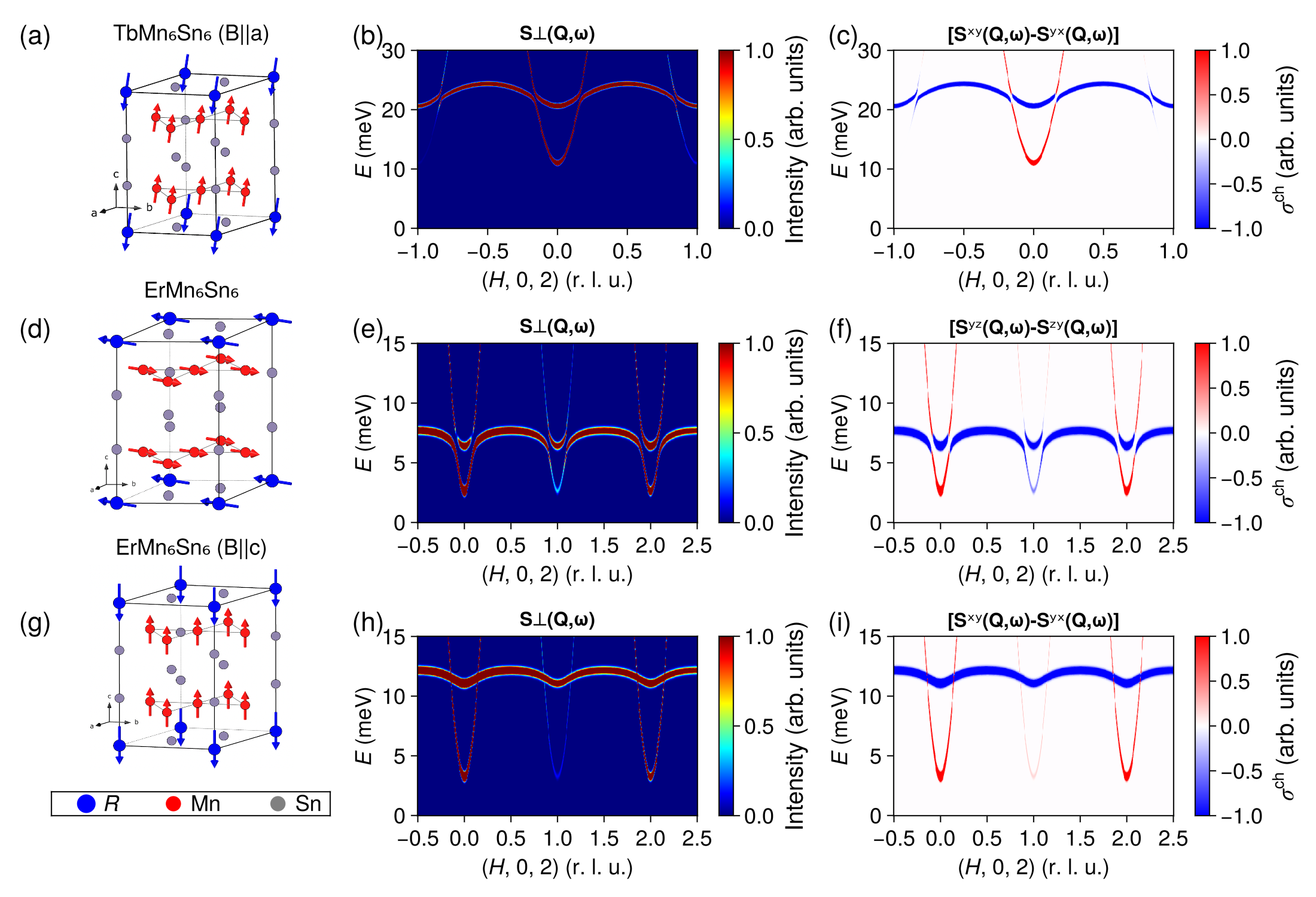}
        \caption{{\bf Switching of hybridized chiral magnons in TbMn$_6$Sn$_6$ and ErMn$_6$Sn$_6$}. (a) Canted ferrimagnetic structure of TbMn$_6$Sn$_6$ via in-plane magnetic field. (b), (c) Unpolarized and chiral scattering spectra in the canted phase of TbMn$_6$Sn$_6$, respectively. (d) Easy-plane ferrimagnetic structure of ErMn$_6$Sn$_6$ at zero magnetic field. (e), (f) Unpolarized and chiral scattering spectra in the $ab-$ plane phase of ErMn$_6$Sn$_6$, respectively. (g) Uniaxial ferrimagnetic structure of ErMn$_6$Sn$_6$ via out-of-plane magnetic field of 1 T. (h), (i) Unpolarized and chiral scattering spectra in the uniaxial phase of ErMn$_6$Sn$_6$, respectively.}
        \label{fig:R166_chi_SI}
    \end{figure}

    $R$166 compounds provide an opportunity to tune magnetic ordering just by applying a magnetic field, changing temperature, or changing rare-earth anisotropy~\cite{Nil25}. For example, Tb166 can be driven from a high-symmetry easy-axis to a low-symmetry easy-plane ferrimagnetic state through the spin-reorientation temperature. Applying an in-plane magnetic field causes canting away from the c-axis, breaking the high-symmetry ferrimagnetic state as shown in Fig.\ref{fig:R166_chi_SI}(a). The calculated unpolarized and chiral scattering spectra are shown in Fig.\ref{fig:R166_chi_SI}(b-c). The unhybridized magnon modes now hybridize, opening a hybridization gap. On the other hand, a small magnetic field can drive Er166 from the easy-plane (Fig.~\ref{fig:R166_chi_SI}(d)) to the easy-axis ferrimagnetic state. Uniaxial magnetic state is shown in Fig.\ref{fig:R166_chi_SI}(g). The unpolarized and chiral scattering spectra of Er166 are shown in Fig.\ref{fig:R166_chi_SI}(h-i). The magnon modes cross over without hybridization or mixing, as in uniaxial ferrimagnetic Tb166. In Er166, it is also possible that magnon hybridization can be tuned by tuning Er anisotropy, while keeping the low symmetry ab-plane ferrimagnetic state. We note that artificially including the $B_6^6$ term in the spin Hamiltonian for Tb166 does not generate any observable mode hybridization.

\begin{figure}[!h]
        \centering
        \includegraphics[width=1.0\linewidth]{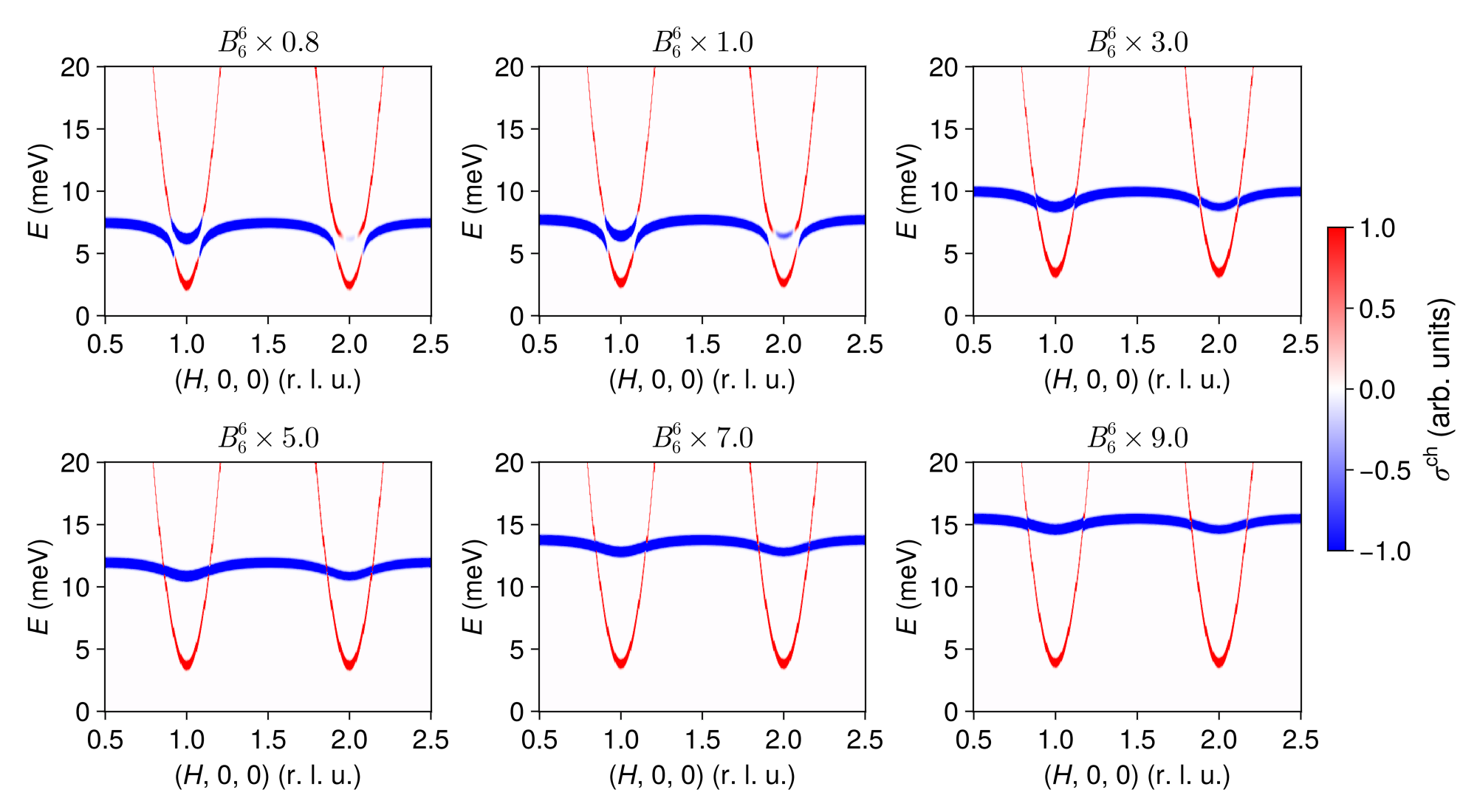}
        \caption{{\bf Hybridization gap dependence on $B_{6}^{6}$ in ErMn$_6$Sn$_6$.} Chiral scattering as a function of $B^6_6$ parameter shows that the hybridization gap decreases continuously with increasing $B_{6}^{6}$ and finally acoustic and optical modes start to cross over without hybridization when $B_{6}^{6}$ is increased by four times. On the other hand, the hybridization gap widens as the $B_4^0$ parameter increases.}
        \label{fig:R166_B66}
    \end{figure}

    Figure~\ref{fig:R166_B66} shows the hybridization gap dependence on the $B_6^6$ parameter in Er166. The hybridization gap continuously decreases as the $B_6^6$ parameter increases, and previously hybridized acoustic and optical modes crossover without hybridization when $B_6^6$ is increased by 5 times its actual value. As shown in Table \ref{tab:aniso}, the classical single-ion anisotropy for Er $K^R_{yy}-K^R_{zz}$ becomes zero (isotropic) in a range of $B_6^6$, and it displays circular precession around its local magnetization. Further increasing the $B_6^6$ parameter, the anisotropy again becomes anisotropic, and modes start to hybridize and open up a gap before it drives magnetic ordering to the c-axis ferrimagnet. The classical single-ion magnetic anisotropy energy (MAE) is discussed in terms of CEF parameters in the following. 

    The classical MAE can be obtained from the $B^l_m$ parameters. For the rare-earth ion in $R$166 compounds, the MAE is given by
    \begin{align}
        E_A = K_1\sin^2{\theta}+K_2\sin^4{\theta}+K_3\sin^6{\theta}+K'_3\cos{(6\varphi)}\sin^6{\theta}
    \end{align}
    where $\theta$ and $\varphi$ are the polar and azimuthal angles of the $R$ moment. The MAE constants are related to the CEF parameters as follows.
    \begin{align*}
        K_1&=-3J^{(2)}B^0_2-40J^{(4)}B^0_4-168J^{(6)}B^0_6 \\
        K_2&=35J^{(4)}B^0_4+378J^{(6)}B^0_6 \\
        K_3&=-231J^{(6)}B^0_6\\
        K'_3&=J^{(6)}B^6_6
    \end{align*}
    where $J^{(2)}=J(J-\frac{1}{2})$, $J^{(4)}=J^{(2)}(J-1)(J-\frac{3}{2})$, and $J^{(6)}=J^{(4)}(J-2)(J-\frac{5}{2})$. The easy-plane configuration with $\theta$=$\frac{\pi}{2}$ is obtained subject to the condition that; $K_1+K_2+(K_3-|K'_3|)< 0$. For a small perturbation around the equilibrium direction, the anisotropy energy to second-order in $\theta$ and $\varphi$ can be expressed as 
    \begin{align*}
        E_A\approx-[K_1+2K_2+3(K_3-|K'_3)|]\theta^2+18|K'_3|\varphi^2
    \end{align*}
    We can associate classical $E_A$ with the single-ion anisotropy term $\mathcal{H}_\text{aniso}=\sum\limits_{\alpha\beta}K^R_{\alpha\beta} J_\alpha\cdot J_\beta\approx K^R_\text{yy}J^2\varphi^2+K^R_\text{zz}J^2\theta^2$. This gives
    \begin{align*}
        K^R_\text{yy}&=18|K'_3|/J^2\\
        K^R_\text{zz}&=-[K_1+2K_2+3(K_3-|K'_3)|/J^2
    \end{align*}
    where $K^R_\text{yy}>0$ and $K^R_\text{zz}>0$ for the planar easy-axis case.
    \begin{table}[!h]
        \centering
        \renewcommand{\arraystretch}{1.5}
        \setlength{\tabcolsep}{7pt} 
        \caption{Classical anisotropy for rare-earth ion Er ($J=15/2$) derived from $B^m_l$ (in meV).}
        \label{tab:aniso}
        \begin{tabular}{l|c|c|c|c|c|c}
            \hline\hline
            & $B^6_6\times0.8$ & $B^6_6\times1.0$ & $B^6_6\times3.0$ & $B^6_6\times5.0$ & $B^6_6\times7.0$ & $B^6_6\times9.0$ \\
            \hline
            $K^R_\text{yy}$ & $0.0848$ & $0.1059$ & $0.3178$ & $0.5297$ & $0.7416$ & $0.9535$ \\
            $K^R_\text{zz}$ & $0.4910$ & $0.4945$ & $0.5298$ & $0.5651$ & $0.6004$ & $0.6358$ \\
            \hline\hline
        \end{tabular}
    \end{table}
\subsection{Magnon mode dynamical chirality}
    \begin{figure}[!h]
        \centering
        \includegraphics[width=0.45\linewidth]{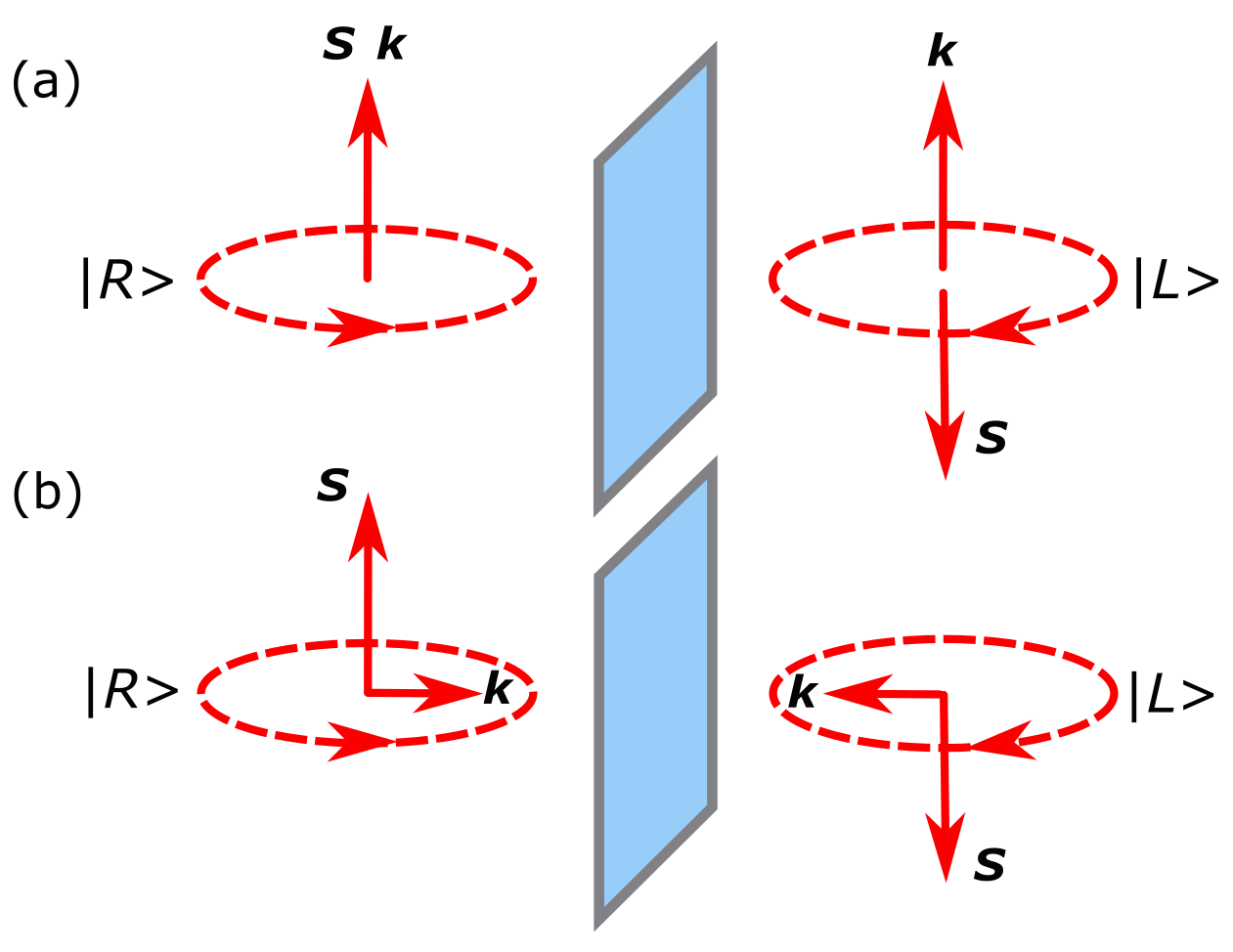}
        \caption{{\bf Magnon mode chirality}. Magnon modes that break improper rotational symmetry are true chiral. (a) A magnon propagating perpendicular to its plane of precession, reverses the motion with respect to the direction of propagation under mirror reflection, whereas remains invariant under time reversal ($\mathcal{M}\boldsymbol{\mathcal{S}}\cdot \bm{k}=-\boldsymbol{\mathcal{S}} \cdot \bm{k}$ and $\mathcal{T}\boldsymbol{\mathcal{S}} \cdot \bm{k}=\boldsymbol{\mathcal{S}}\cdot \bm{k}$). (b) A magnon propagating parallel to its plane of precession reverses its motion as well as the direction of propagation under mirror reflection, so it falls into false chirality ($\bm{S}\cdot \bm{k}=0$). True chiral enantiomers can be interconverted by a reflection, but not by time reversal: $|L> =\mathcal{M}|R>\neq \mathcal{T}|R>$. However, false chiral enantiomers can be interconverted by either a reflection or by time reversal: $|L> =\mathcal{M}|R>=\mathcal{T}|R>$.}
        \label{fig:mag_chi}
    \end{figure}

\bibliography{reference.bib}